\input harvmac

\let\includefigures=\iftrue
\newfam\black
\includefigures
\input epsf
\def\figin{\epsfcheck\figin}\def\figins{\epsfcheck\figins}
\def\epsfcheck{\ifx\epsfbox\UnDeFiNeD
\message{(NO epsf.tex, FIGURES WILL BE IGNORED)}
\gdef\figin##1{\vskip2in}\gdef\figins##1{\hskip.5in}
\else\message{(FIGURES WILL BE INCLUDED)}%
\gdef\figin##1{##1}\gdef\figins##1{##1}\fi}
\def\DefWarn#1{}
\def\figinsert{\goodbreak\midinsert}
\def\ifig#1#2#3{\DefWarn#1\xdef#1{fig.~\the\figno}
\writedef{#1\leftbracket fig.\noexpand~\the\figno}%
\figinsert\figin{\centerline{#3}}\medskip\centerline{\vbox{\baselineskip12pt
\advance\hsize by -1truein\noindent\footnotefont{\bf Fig.~\the\figno:} #2}}
\bigskip\endinsert\global\advance\figno by1}
\else
\def\ifig#1#2#3{\xdef#1{fig.~\the\figno}
\writedef{#1\leftbracket fig.\noexpand~\the\figno}%
\global\advance\figno by1}
\fi


\def\np#1#2#3{Nucl. Phys. {\bf{B{#1}}} (#2) #3}
\def\pl#1#2#3{Phys. Lett. {\bf{B{#1}}} (#2) #3}

\def\prl#1#2#3{Phys. Rev. Lett. {\bf{{#1}}} (#2) #3}
\def\cmp#1#2#3{Comm. Math. Phys. {\bf{{#1}}} (#2) #3}
\def\mpl#1#2#3{Mod. Phys. Lett. {\bf{A{#1}}} (#2) #3}
\def\jhp#1#2#3{JHEP~{\bf{#1}} ({#2}) {#3}}

\def\cqg#1#2#3{Class. Quant. Grav. {\bf{#1}} (#2) {#3}}


\def\tq{{\tilde{q}}}
\def\tp{{\tilde{\psi}}}
\def\tx{{\tilde{x}}}
\def\tal{{\tilde{\alpha}}}

\def\dr{\rangle\!\rangle}

\def\ca{{\cal A}}
\def\ct{{\cal T}}
\def\ha{{1\over{2}}}
\def\coherentr#1#2#3{|D{#1};{#2};{#3}\dr}

\def\bc{{\bf C}}
\def\vx{{\bf x}}
\def\bv{{\bf v}}
\def\bw{{\bf w}}
\def\bp{{\bf p}}
\def\bz{{\bf Z}}


\nref\DM{M.~R.~Douglas and G.~Moore, {\it D-Branes, Quivers, and ALE
Instantons}, hep-th/9603167.}

\nref\fuchs{J.~Fuchs and C.~Schweigert,
{\it Branes: From Free Fields to General Backgrounds},
Nucl. Phys. {\bf B530}, 99 (1998)
hep-th/9712257.}

\nref\RS{A.~Recknagel and V.~Schomerus, {\it D-branes in Gepner
Models}, \np{531}{1998}{185}, hep-th/9712186.}

\nref\GSone{M.~Gutperle and Y.~Satoh, {\it D-Branes in Gepner Models
and Supersymmetry}, \np{543}{1999}{73}.}

\nref\GStwo{M.~Gutperle and Y.~Satoh, {\it D0-Branes in Gepner Models
and $N=2$ Black Holes}, hep-th/9902120.}

\nref\RW{S.~Ramgoolam and D.~Waldram, {\it Zero-Branes on a Compact
Orbifold}, \jhp{9807}{1998}{009}, hep-th/9805191.}

\nref\GLY{B.~Greene, C.~I.~Lazaroiu and P.~Yi, {\it D-Particles on
$T^4/\bz_N$ and their Resolutions}, \np{539}{1999}{135}, hep-th/9807040.}

\nref\hori{K.~Hori, {\it D-Branes, T-Duality, and Index Theory},
hep-th/9902102.}

\nref\JLC{J.~L.~Cardy, {\it Boundary Conditions, Fusion Rules and
the Verlinde Formula}, \np{324}{1989}{581}.}

\nref\ISH{N.~Ishibashi, {\it The Boundary and Crosscap States in
Conformal Field Theories}, \mpl{4}{1989}{251}.} 

\nref\BGsdual{O.~Bergman and M.~R.~Gaberdiel, {\it A
Non-Supersymmetric Open String Theory and S-Duality},
\np{499}{1997}{183},
hep-th/9701137.}

\nref\SENtwo{A.~Sen, {\it Stable non-BPS Bound States of BPS D-branes}
\jhp{9808}{1998}{010}, hep-th/980519.}

\nref\CLNYtwo{C.~G.~Callan, C.~Lovelace, C.~R.~Nappi and S.~A.~Yost, {\it
Adding Holes and Crosscaps to the Superstrings}, \np{293}{1987}{83}.}

\nref\PC{J.~Polchinski and Y.~Cai, {\it Consistency of Open
Superstring Theories}, \np{296}{1988}{91}.}

\nref\GG{M.~B.~Green and M.~Gutperle, {\it Light Cone Supersymmetry
and D-Branes}, \np{476}{1996}{484}, hep-th/9604091.}

\nref\divec{P.~Di~Vecchia, M.~Frau, I.~Pesando, S.~Sciuto,
A.~Lerda and R.~Russo, {\it Classical p-Branes from Boundary States},
\np{507}{1997}{259}, hep-th/9707068.}

\nref\HINS{F.~Hussain, R.~Iengo, C.~Nunez and C.~Scrucca,
{\it Interaction of Moving D-branes on Orbifolds},
\pl{409}{1997}{101}, hep-th/9706186.}

\nref\BCR{M.~Billo, B.~Craps and F.~Roose,
{\it On D-branes in Type 0 String Theory},
hep-th/9902196.}

\nref\BGone{O.~Bergman and M.~R.~Gaberdiel, {\it Stable non-BPS
D-particles}, \pl{441}{1998}{133}, hep-th/9806155.}

\nref\AO{M.~Oshikawa and I.~Affleck, {\it Boundary Conformal Field
Theory Approach to the Critical Two-Dimensional Ising Model with a
Defect Line}, \np{495}{1997}{533}, cond-mat/9612187.}

\nref\BGtwo{O.~Bergman and M.~R.~Gaberdiel, {\it Non-BPS States in
Heterotic Type IIA Duality}, \jhp{9903}{1999}{013}, hep-th/9901014.}

\nref\DDG{D.-E.~Diaconsecu, M.~R.~Douglas and J.~Gomis, {\it
Fractional Branes and Wrapped Branes}, \jhp{9802}{1998}{013},
hep-th/9712230.}

\nref\LMN{J.~Lauer, J.~Mas and H.~P.~Nilles, {\it Duality and the
Role of Non-perturbative Effects on the World Sheet},
\prl{226}{1989}{251}.}

\nref\DFMS{L.~Dixon, D.~Friedan, E.~Martinec and S.~Shenker, {\it The
Conformal Field Theory of Orbifolds},
\np{B282}{1987}{13}.}

\nref\HV{S.~Hamidi and C.~Vafa, {\it Interactions on Orbifolds},
\np{B279}{1987}{465}.}

\nref\DVV{R.~Dijkgraaf, E.~Verlinde and H.~Verlinde, {\it Conformal
Field Theories on Riemann Surfaces}, \cmp{115}{1988}{649}.}

\nref\GPR{A.~Giveon, M.~Porrati and E.~Rabinovici, {\it Target Space
Duality in String Theory}, Phys. Rep. {\bf244} (1994) 77,
hep-th/9401139.}

\nref\LLW{W.~Lerche, D.~L\"ust and N.~Warner, {\it Duality Symmetries
in $N=2$ Landau-Ginzburg Models}, \pl{231}{1989}{417}.}

\nref\ASPkthree{P.~S.~Aspinwall, {\it $K3$ Surfaces and String Duality},
in TASI 1996, {\it Fields, Strings and Duality}, hep-th/9611137.}

\nref\ASPpoint{P.~S.~Aspinwall and D.~R.~Morrison,
{\it Point-like Instantons on $K3$ orbifolds},
\np{503}{1997}{533}, hep-th/9705104.}

\nref\BSV{M.~Bershadsky, V.~Sadov and C.~Vafa, {\it D-branes and
Topological Field Theories}, \np{463}{1996}{420}, hep-th/9511222.}

\nref\GHM{M.~B.~Green, J.~Harvey and G.~Moore, {\it I-brane Inflow and
Anomalous Couplings on D-branes}, \cqg{14}{1997}{47}, hep-th/9605033.}


\Title{\vbox{\baselineskip12pt\hbox{hep-th/9905078}
\hbox{RU-99-21}}}
{\vbox{
\centerline{D-branes on $T^4/$Z$_2$ and T-Duality} }}
\centerline{Ilka Brunner, Rami Entin and Christian R\"omelsberger}
\medskip
\centerline{\it Department of Physics and Astronomy}
\centerline{\it Rutgers University }
\centerline{\it Piscataway, NJ 08855--0849}
\medskip
\centerline{\tt ibrunner,rami,roemel@physics.rutgers.edu}
\medskip
\bigskip
\noindent
In this note we discuss D-branes on $T^4/\bz_2$ using the boundary
states formalism. Explicit formulas for the untwisted boundary states
inherited from the underlying $T^4$ and twisted states corresponding
to branes wrapping collapsed $2$-cycles at the orbifold singularities
are given. The exact CFT description of the orbifold makes it possible
to study how the boundary states transform under
$R_i\mapsto {1\over{R_i}}$ transformation on all directions of the
underlying $T^4$.  We compare their transformation law with results
obtained from world volume considerations.

\Date{May 1999}

\newsec{Introduction}

Starting in \DM, D-branes transverse to $\bc^2/\bz_N$ orbifolds
have been discussed extensively in the literature. Much less has been
said about D-branes moving in curved compact backgrounds and on
compact orbifold backgrounds, or even general CFT backgrounds. (See
however \fuchs\ and \refs{\RS,\GSone,\GStwo} which discuss D-branes in Gepner
models.) In this paper,
we will study the toroidal orbifold $T^4/\bz_2$. This model
has been studied from the point of view of world volume
theories of D-branes in \refs{\RW,\GLY,\hori}. Here,  a
different point of view will be taken, namely, we will construct
boundary states for the D-branes. This approach goes back to
the work of Cardy in the context of rational conformal field
theory \JLC. Let us briefly describe the main ideas
of this work. 

The CFT description of open strings involves
Riemann surfaces with boundaries as world sheets, e.g. the strip
or the upper half plane. In the interior, the theory behaves
like a theory on the full plane. The physics on the boundary
depends on the boundary conditions imposed there.
Consider a bulk theory realizing two isomorphic left and right-moving
chiral algebras. On the boundary the left and
right-moving generators must be related by an automorphism of the
chiral algebra, otherwise the corresponding symmetry will not be
preserved. In particular, since no energy is allowed to flow across
the boundary, the operators  $T(z)$ and $\tilde T(\bar z)$ must be
identified on the boundary. 

These boundary conditions
can be transformed to the closed string sector. In this
language, the boundary conditions are implemented on
boundary states, as given in \ISH. However, for a consistent
boundary state modular invariance imposes additional conditions.
To see this, consider an open string one-loop amplitude, i.e.
the world sheet is a cylinder. Alternatively, this diagram
can be viewed as a closed string tree-level amplitude between
boundary states. This leads to the condition, that the modular
transformation of a transition amplitude between boundary states
leads to an open string partition function, in particular,
a sum of characters with integer coefficients (Cardy's condition).

The plan of this paper is as follows. In the next section we establish
our conventions and notations by reviewing supersymmetric boundary
states on tori. Our discussion will be brief and the reader is
referred to \refs{\BGsdual,\SENtwo} for a more complete treatment. The
twisted and untwisted orbifold boundary states are then constructed in
section 3 where we also discuss their spacetime interpretation. 

In section  4, we discuss T-duality on a
$T^4/\bz_2$ orbifold. More precisely, we consider a transformation
$\ct$, which acts on the untwisted sector of the orbifold theory as the
$R_i\mapsto{1\over{R_i}}$ transformation along all four
circles of the underlying $T^4$. We point out that if $\ct$ acts on
the twist fields as a certain unitary transformation, it preserves 
the 3-point functions and is therefore a symmetry
of the CFT. The action of $\ct$ on the boundary states is then
compared with a transformation given in \RW\ acting on the BPS vectors
of the orbifold limit of type~IIA on $K3$. We find agreement
between the two. In the last section we discuss our results.

\newsec{Boundary states in toroidal compactifications}

Let us briefly review the construction of boundary states
\refs{\CLNYtwo,\PC} in a toroidal background where $x^6,\ldots,x^9$
are compactified with radii $R_6,\ldots,R_9$. To avoid introducing
(super)ghosts, we will work  in the light-cone gauge as in \GG\ and
consider Dirichlet boundary conditions localizing the D-branes at the
origin of all non-compact directions. The light-cone boundary
conditions mean that we we will effectively describe $p+1$-instantons
instead of $p$-branes, but these are related by a double Wick
rotation. The problem at hand is constructing the NSNS and RR sector
boundary states $\coherentr{p}{\eta}{\vec k}_{NSNS\atop{RR}}$
satisfying the boundary conditions
\eqn\bcone{\eqalign{
(\alpha^i_n-\tal^i_{-n})\coherentr{p}{\eta}{\bf k}_{NSNS\atop{RR}}&=0\cr
(\alpha^\mu_n+\tal^\mu_{-n})\coherentr{p}{\eta}{\bf k}_{NSNS\atop{RR}}&=0\cr
(\psi_r^i-i\eta\tp_{-r}^i)\coherentr{p}{\eta}{\bf k}_{_{NSNS}}&=0\cr
(\psi_r^\mu+i\eta\tp_{-r}^\mu)\coherentr{p}{\eta}{\bf k}_{_{NSNS}}&=0\cr 
(\psi_r^i-i\eta\tp_{-r}^i)\coherentr{p}{\eta}{\bf k}_{_{RR}}&=0\cr
(\psi_r^\mu+i\eta\tp_{-r}^\mu)\coherentr{p}{\eta}{\bf k}_{_{RR}}&=0}
\hskip 40pt\eqalign{&n\in\bz\cr\cr&r\in{\bf
Z}+\half\cr\cr&r\in\bz}
}
where $\mu,\nu$  the Neumann directions and $i,j$ label the Dirichlet
ones. 
The vector ${\bf k}$ denotes the continuous momenta carried
by the bosonic Fock vacuum in the non-compact directions
$x^2,\ldots,x^5$ and the discrete momentum or winding quantum numbers
in the compact directions $x^6,\ldots,x^9$. The parameter $\eta=\pm1$
reflects the freedom of choosing a spin structure on the cylinder \PC. 

In the bosonic sector the equations \bcone\ are solved by the 
coherent state
\eqn\ishibashi{|Bp;{\bf k}\dr=\exp\left\{\sum_{n>0}{1\over{n}}(\alpha^i_{-n}\tal_{-n}^i-\alpha_{-n}^\mu\tal_{-n}^\mu)\right\}|{\bf k}\rangle.}
In particular, the zero mode equations are  solved in the Dirichlet
directions by vacua $|B;n_i\rangle$ carrying momenta ${n_i\over{R_i}}$
and in the Neumann directions by vacua $|B;m_\mu\rangle$ carrying
winding $2m_\mu R_\mu$. The  states \ishibashi\ are not yet consistent
boundary states in the sense of Cardy, since the tree level amplitudes
between them do not transform under the $S$-modular transformation into
open string partition functions. 

In order to obtain the correct D-brane states it is necessary to localize the
branes in the Dirichlet directions and assign definite Wilson lines
in the Neumann directions. This is done by Fourier transforming with respect
to the positions $x_i$ and the Wilson lines $\tx_\mu$. The correct
boundary state for a D$p$-brane sitting at the origin in the external
dimensions is therefore given by
\eqn\bsone{\eqalign{
|Bp;\vx\rangle&=\int\prod_{l=0}^5dk_l|B;k_l\dr_D\cr
&\times\prod_{j=6}^{6+p}{1\over{\sqrt{2R_j}}}\sum_{n_j\in\bz}e^{-in_jx_j\over R_j}|B;n_j\dr_D
\prod_{\mu=7+p}^{9}{\sqrt{R_\mu}}\sum_{m_\mu\in\bz}e^{-i2R_\mu m_\mu\tx_\mu}|B;m_\mu\dr_N,}} 
where $\vx$ encodes both $x_i$ and $\tx_\mu$. For completeness and
later use, the tree level transition amplitude between two parallel
$p$-branes is then given by
\eqn\transampone{\eqalign{
\ca^B_{p-p}(\Delta\vx)&=\int_0^\infty{d\tau\over{\tau^2}}\ca^B_{p-p}(\Delta\vx;\tq)
=\int_0^\infty{d\tau\over{\tau^2}}\int\!\prod_{l=0}^5dk_l{e^{-{\pi\over{\tau}}k_l^2}\over{\eta(\tq)^4}}\cr
&\cdot{1\over{\eta(\tq)^4}}\prod_{j=6}^{6+p}{1\over{2R_j}}\sum_{n_j\in{\bf
Z}}e^{in_j\Delta x_j\over R_j}\tq^{({n_j\over{2R_j}})^2}
\cdot\prod_{\mu=7+p}^{9}R_\mu\sum_{m_\mu\in\bz}e^{2im_\mu
R_\mu\Delta\tx_\mu}\tq^{(2m_\mu R_\mu)^2}
}}
The calculation of the corresponding open string amplitude function is
standard and yields
\eqn\opentwo{\eqalign{
{\cal Z}_{op}^B(\Delta\vx)=&\int_0^\infty{d\tau\over{\tau}}{\cal Z}^B_{op}(q;\Delta\vx)
=\int_0^\infty{d\tau\over{\tau}}{1\over{\eta(q)^4}}\cr
\cdot&{1\over{\eta(q)^4}}\prod_{j=p}^{6+p}\sum_{n_j\in\bz}q^{\half(2R_jn_j+{\Delta x_j\over{\pi}})^2}\cdot \prod_{\mu=7+p}^{9}\sum_{m_\mu\in\bz}q^{\half({m_\mu\over{R_\mu}}+{\Delta\tx_\mu\over{\pi}})^2} 
.}} 
It can be checked that \transampone\ and \opentwo\ are related by
$S$-modular transformation as required. Observe that the difference
$\Delta\vx$ the parameters encoding the positions and Wilson lines
leads to a shift of the ground state energy of the different open
string sectors as expected. 

As in the bosonic case, the fermionic conditions in \bcone\ are solved
by coherent fermionic states
\eqn\fermi{|Fp;\eta\dr_{_{NSNS\atop{RR}}}=\exp\left\{i\eta\sum_{r>0}\left(\psi^i_{-r}\tp^i_{-r}-\psi^\mu_{-r}\tp^\mu_{-r}\right)\right\}|F_{g.s.};\eta\rangle_{_{NSNS\atop{RR}}}.}
In the NSNS sector $|F_{g.s.}\rangle_{_{NSNS}}=|0\rangle$ is the
unique ground state and has no $\eta$ dependence. In the RR sector,
$|F_{g.s.}\rangle_{_{RR}}$ must also solve the zero mode equations
in \bcone, and therefore depends non-trivially on $\eta$. The
different zero mode solutions encode $p$-brane of all dimension
$p=1,\ldots,4$ as indicated in \bcone\ \divec.

The action of the fermionic number operator on the different states 
\fermi\ is given by
\eqn\susyone{\eqalign{
&(-1)^F|Fp;\eta\rangle_{_{NSNS}}=-|Fp;-\eta\rangle_{_{NSNS}}
\hskip30pt(-1)^{\tilde F}|Fp;\eta\rangle_{_{NSNS}}=-|Fp;-\eta\rangle_{_{NSNS}}\cr
&(-1)^F|Fp;\eta\rangle_{_{RR}}=(-1)^{7-p}|Fp;-\eta\rangle_{_{RR}}\hskip24pt(-1)^{\tilde F}|Fp;\eta\rangle_{_{RR}}=|Fp;-\eta\rangle_{_{RR}}
}}
It follows that the GSO invariant boundary states are
\eqn\susytwo{\eqalign{
|Fp\rangle_{_{NSNS}}&={1\over{\sqrt{2}}}\left(|Fp;+\rangle_{_{NSNS}}-|Fp;-\rangle_{_{NSNS}}\right)\cr
|Fp\rangle_{_{RR}}&={1\over{\sqrt{2}}}\left(|Fp;+\rangle_{_{RR}}+|Fp;-\rangle_{_{RR}}\right),
}}
where $p$ is even in type~IIA theory and odd in type~IIB. 

By world sheet duality, the transition amplitude between the GSO
projected boundary states reproduces the NS and R sectors of the open
string partition function. The same reasoning also shows that the
transition amplitude between two (NSNS) RR boundary states becomes the
open string partition function with (no) $(-1)^F$ inserted. An explicit
calculation gives the correct supersymmetric $p$-brane state as
\eqn\susythree{|Dp;\vx\rangle=|Bp;\vx\rangle\otimes{1\over{\sqrt{2}}}\left(|Fp\rangle_{_{NSNS}}\pm4i|Fp\rangle_{_{RR}}\right).}
The RR charges of a D-brane are determined by saturating the boundary
state with the corresponding RR vertex operator as in \divec. By
convention, the $+$ sign above is taken to be a brane and the $-$
sign to an anti-brane.

\newsec{D$p$-branes on the $\bz_2$ orbifold}

In this section D-branes on toroidal orbifolds are studied
in the boundary
state language. Previous work in a similar direction has 
appeared in \refs{\HINS, \BCR} see also \refs{\SENtwo, \BGone}
for a discussion in the context of non-BPS but stable
states.

From open string considerations it is well known that there
are different types of D-branes on orbifolds, depending on the
representation of the orbifold group chosen on the Chan-Paton factors
\refs{\DM,\DDG}. On a $\bz_2$ orbifold there is the regular representation 
and one-dimensional representations.
The resulting D-branes differ physically in their RR-charges:
While the regular representation leads to a brane which is
only charged under untwisted sector fields, the one-dimensional representations
give D-branes which are charged under both twisted and
untwisted sector RR-fields. In the boundary state language this means
that the former type of brane contains only Ishibashi
states built over untwisted sector primary fields, whereas the
latter contains an Ishibashi state built over a twist field.
Therefore, we refer to them as ``untwisted'' and ``twisted''
branes respectively. In this section we will discuss both
untwisted and twisted D$0$-branes in detail, whereas part of the
discussion of the higher dimensional branes is postponed to
a later section.

\subsec{The untwisted states}

The untwisted boundary states are $\bz_2$ invariant
combinations of the $T^4$ boundary states.
The orbifold action reverses the sign of both the fermionic
and bosonic oscillators in the $a=6,\ldots,9$ directions:
\eqn\orbac{
\alpha_{n}^{a} \to -\alpha_{n}^{a}, \quad \psi_r^{a} \to 
-\psi_r^{a},}
and the action on the RR vacua given by
\eqn\orbrrac{
|F_{g.s.}\rangle_{_{RR}} \to \prod_{a=6}^9\sqrt{2}\psi_0^a\prod_{a=6}^9\sqrt{2}\tp_0^a
|F_{g.s.}\rangle_{_{RR}}.}
The RR vacua, on which the even dimensional branes are built, are invariant
under this action as can be verified  using the conditions
\bcone. Moreover, odd dimensional BPS branes in the orbifold directions
are projected out as required by the homology of a $K3$ surface. Using
this, we see that the full boundary state transforms as
\eqn\orbact{|Dp;\vx \rangle \to |Dp; -\vx \rangle,}
leading to the invariant states
\eqn\orbstateone{|Dp;\vx\rangle^{Orb}={1\over{\sqrt2}}\left(|Dp;\vx\rangle+|Dp;-\vx\rangle\right).
}
The states \orbstateone\ are built only over untwisted vacua and for this
reason are charged under the untwisted RR fields as the $T^4$ states.
The computation of the transition amplitudes between two such boundary
states is reduced to a computation in the unorbifolded theory. For
example the transition amplitude of a boundary state with itself is
\eqn\orbipart{\eqalign{
\langle Dp;\vx |\tilde q^{\ha H_{cl}} | Dp; \vx
\rangle^{orb}
&={1\over 2}\langle Dp;\vx|\tilde q^{\ha H_{cl}} | Dp; \vx
\rangle^{torus} 
+{1\over 2}\langle Dp;-\vx|\tilde q^{\ha H_{cl}} | Dp; -\vx
\rangle^{torus}\cr
&+ {1\over 2}
\langle Dp;\vx|\tilde q^{\ha H_{cl}} | Dp; -\vx
\rangle^{torus} + {1\over 2}
\langle Dp;-\vx|\tilde q^{\ha H_{cl}} | Dp; \vx
\rangle^{torus}.
}}
The relevant amplitudes for the torus were given in the previous
section. As expected, the $S$-modular transformation of this result
leads to an open string partition function with the regular
representation on the Chan-Paton factors. The first two terms in this
sum are equal to each other and so are the last two. They correspond
respectively to strings stretching between a D-brane and itself and
strings stretching between a D-brane and its image. If the D-branes
are away from the fixed points \foot{The term ``fixed point''
has been used somewhat loosely, since  the parameters
for the Neumann directions are Wilson lines.},
only the first two terms can lead to marginal operators, with the
corresponding moduli being the position and Wilson line of the brane.
At the fixed point $\vx=-\vx$, there are additional marginal
operators coming from the last two terms in the sum \orbipart.
These arise from external directions to the orbifold, and in the
D$0$-brane world volume theory signal a Coulomb branch opening
\refs{\DM, \DDG}. The new massless fields arising from the
internal directions are gauge equivalent to the marginal operators
already in the theory.

The BPS property and the RR charges carried by \orbstateone\ are
crucial in correctly identifying these states with
spacetime D-branes. The state with $p=0$ is a D$0$-brane which can
move anywhere inside the orbifold. It carries one unit of D$0$-brane
charge, and has four position moduli. The state with $p=2$ corresponds
to a D$2$-brane wrapping once one of the six $T^2$ cycles inherited
from the underlying $T^4$. This assignment again fixes the absolute
normalization of the D$2$-brane charge. Their moduli are equally well
understood as transverse position and Wilson line. This can be seen by
regarding  $T^4/\bz_2$
as an elliptic fibration over a base ${\bf P}^1$ with four $D_4$
singular fibers. The inherited $2$-cycles which the D$2$-branes wrap are
elliptic curves with respect to a properly chosen complex structure,
and have one complex deformation. These four moduli account
respectively for the two $\tx_\mu$ and two $x_i$ parameters which
appear in the boundary state.
The state with $p=4$ is a state carrying only D$4$-brane charge
wrapping the orbifold. We will discuss its physics in the section
on T-duality.

\subsec{The twisted case}

Let us now turn to the twisted boundary states,
generalizing the discussion of the bosonic $S^1/\bz_2$ orbifold in \AO. 
The 16 fixed points  of the orbifold are located on points in
${1\over2}W$, where $W$ denotes the winding lattice. Therefore
it is natural to label the bosonic twist fields $T_{[\bv]}$ by cosets
$[\bv]$, where $\bv \in {1\over 2}W$.
Each of the 16 twisted vacua $|T_{[\bv]}\rangle$
produces an all Dirichlet coherent state 
\eqn\twistboseone{|DT_{[\bv]}\dr=\exp\{\sum_{n={1\over 2}}^\infty
{1\over{n}}\alpha^i_{-n}\tal^i_{-n}\}|T_{[\bv]}\rangle.}
Taking into account the fermions, we note that in the twisted NSNS
(RR) sector the moding of the fermions is (half)
integral, and hence there is a unique twisted RR ground state.
However, there are fermionic zero modes in the NSNS orbifold directions.
The twisted NSNS ground state, which fulfills the zero-mode part of \bcone\
must be picked by the same considerations as for the RR ground state
on the torus.

The RR Ishibashi state leads to a charge under
the corresponding RR vertex operator, implying that the Ishibashi
state should correspond to D$2$-brane wrapped on a collapsed sphere at
a fixed point. It is known from the gauge theory analysis that these
D$2$-branes also carry one half unit of D$0$-brane charge induced by 
the $B$-field on the collapsed cycles. 
In CFT language this is reflected in the requirement of modular
invariance. Let us consider open strings starting and ending at the
same fixed point with Chan Paton factors transforming in the trivial
representation of the $\bz_2$ orbifold group. The open string
partition function with the $\bz_2$ generator $g$ inserted reproduces the
amplitude between a twisted Ishibashi state and itself. The other
part of the open string partition function gives a D$0$-brane
amplitude. The coefficients of the twisted and untwisted parts are
fixed by the bosonic open string partition function given by
\eqn\orbone{\eqalign{
{\cal Z}_{op}^B(q)&={\rm Tr}_{_{{\cal H}_{op}}}{1+g\over{2}}q^{H^B_{op}}\cr
&=\ha{\cal Z}^B_{op}(q;0)+\ha\left(q^{-{1\over{24}}}\prod_{n=1}^\infty{1\over{1+q^n}}\right)^4
=\ha{\cal Z}^B_{op}(q;0)+\ha\left(\sqrt{\eta(q)\over{\theta_2(q)}}\right)^4.
}}
This shows that in the purely bosonic case the correct fractional
D$0$-brane boundary states are given by
\eqn\orbstatetwo{|B0B2_{[\bv]}\rangle_\pm=2^{-\ha}|B0;{\bf v}\rangle
\pm2^\ha|{BT_{[\bv]}}\dr.}
It can be checked that these states are mutually consistent among
themselves. Note that position of the untwisted part is determined
by the twisted part. All other combinations are ruled
out by consistency with twisted D$4$-brane states which we will
discuss later. In the full supersymmetric theory there are a-priori
eight different states associated to each of the fixed points
corresponding to eight possible relative signs. However, only
the 
\eqn\twistedstates{
|D0D2_{[\bv]}\rangle_{+}, \quad  
|\overline{D0}\overline{D2}_{[\bv]}\rangle_{+}, \quad 
|\overline{D0}D2_{[\bv]}\rangle_{-}, \quad
|D0\overline{D2}_{[\bv]}\rangle_{-}.
}
have a consistent superstring interpretation \BGtwo.

\newsec{T-duality on the orbifold}

\subsec{The T-duality action on the twist fields}

In this section we work out a consistent action of the `inherited
T-duality' transformation $\ct$ on the twist fields of a $T^d/\bz_2$
orbifold. Similar calculations were done in \refs{\DVV,\LMN}.
We take the underlying $T^d$ to be a rectangular torus with radii $R_i$ and
with no $B$-field turned on. As is well known the
spectrum and interactions of the untwisted sector are invariant under
$\ct$, and we therefore limit our discussion to the twisted sector.

Consider first the bosonic sector of a superstring CFT. The highest weight
of a twisted sector is $h_T=\tilde h_T={d\over 16}$. Thus, in order to
preserve the spectrum, the action of $\ct$ can only map twist fields
into a linear combination of twist fields. We will now determine the
exact map by requiring the three point functions involving twist
fields to be invariant. This implies that both the spectrum and
interactions are invariant, so that $\ct$ is a symmetry of the theory.

On the $T^d$ torus with no $B$-field the momentum lattice $M$ and the
winding lattice $W$ decouple. We denote $\bz_2$-invariant combinations 
of untwisted sector vertex operators carrying momentum $\bp$ and winding
$\bw$ in these lattices as $V_{\bp,\bw}$. The  OPE
of two twist fields in a $\bz_2$ orbifold of this torus was given in
\refs{\DFMS,\HV} as follows. The topological selection rule for these
amplitudes are very easily seen from the winding lattice.  

\ifig\fuse{Fusion of a twisted string and a winding string}
{\epsfxsize2.0in\epsfbox{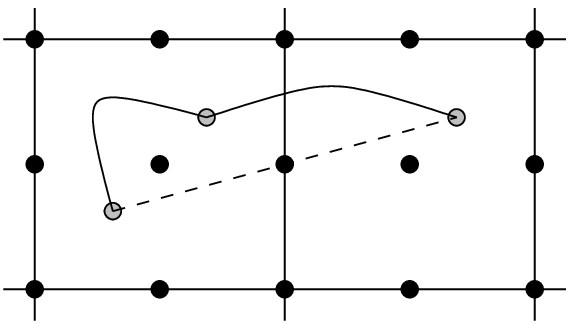}}

The figure above illustrates the joining of a twisted string
$T_{[\bv]}$ with a winding string  $V_{\bp,\bw}$. The result has to be
a twisted string around the fixed point $[\bv+{1\over
2}\bw]$. Factorization of the four point amplitude of twist fields
gives the exact coefficients in the OPE
\eqn\ope{\eqalign{
T_{[\bv_1]}T_{[\bv_2]}=\sum_{{\bp\in M\atop{\bw\in 2W+2\bv_1+2\bv_2}}}
{1\over 2}\left((-1)^{2\bp\cdot\bv_1}+(-1)^{2\bp\cdot\bv_2}\right)
16^{-{1\over 2}(\bp^2+\bw^2)}V_{\bp,\bw}.
}}
The transformation $\ct$ exchanges winding and momentum, and therefore
must also act on the twist fields in order to preserve the OPE
\ope. As we show in the appendix, this is achieved by the transformation 
\eqn\transtwist{
T_{[\bv^\prime]}=2^{-{d\over 2}}\sum_{[\bv]\in{1\over 2}W/W}
(-1)^{4\bv\cdot\bv^\prime}T_{[\bv]}}
and therefore $\ct$ is a symmetry of the bosonic orbifold CFT. Note
that $\bv^\prime\in\ha M$, so that the inner product
$\bv\cdot\bv^\prime$ above is well defined.

The supersymmetric case is restricted to $d=4$ orbifolds by the
$\bz_2$ action, but presents no new subtleties. The
transformation of the twisted NSNS, RR, NSR or RNS world sheet
fermions under $\ct$ is identical to that of the
untwisted RR, NSNS, RNS or NSR fermions respectively. It follows
that $\ct$ with the usual action on the untwisted sector and the
action \transtwist\ on the twisted sector is a symmetry of 
perturbative superstrings compactified on this $T^4/\bz_2$ orbifold.

It will prove convenient later to adopt an alternative numbering scheme
for the twist fields. There is a natural correspondence between the
conjugacy class $[\bv]$ of each of the 16 fixed points and four-digit
binary numbers $b$, such that the origin is assigned to $0000$ and so
forth. We relabel a twist field  $T_{[\bv]}$ by $T_a$, where
$a=16-b$. In this notation the action \transtwist\ of $\ct$ on the
twist fields is represented by the $16\times16$ matrix 
\eqn\tmatrixone{\ct_{a^\prime a}=\left({1\over{\sqrt{2}}}
\pmatrix{-1&1\cr1&1}\right)^{\otimes4}.}

\subsec{The world volume analysis}

In order to compare the previous results with the transformation \RW\
obtained by world volume analysis, let us briefly recall some relevant
facts about type~IIA theory compactified on a $T^4/\bz_2$ limit of
$K3$. The total integral cohomology of $K3$ forms the lattice
$\Gamma_{4,20}$ \ASPkthree, whose basis vectors are associated to spacetime
BPS states. The inner product on this lattice is the oriented intersection
number of the corresponding cycles of $K3$. It is natural to work with
orthogonal decomposition of this lattice
%
$\Gamma_{4,20}=\Gamma_{3,19}\oplus\Gamma_{1,1}$
%
corresponding to
%
$H^*(K3,\bz)=H^2(K3,\bz)\oplus H^0(K3,\bz)\oplus H^4(K3,\bz)$.
%
In this decomposition, the sub-lattice $\Gamma_{1,1}$ is spanned by
vectors $\omega$ and $\omega^*$ generating $H^4(K3,\bz)$ and
$H^0(K3,\bz)$ respectively over the integers; their intersection form
is given by
\eqn\hp{\omega\cdot\omega=\omega^*\cdot\omega^*=0\qquad\omega\cdot\omega^*=1.}
Although there exists a basis for $\Gamma_{3,19}$ generating
$H^2(K3,\bz)$ over the integers, in the orbifold limit it proves more
convenient to work with a different one. The
elements of this basis are $\omega_i$, $\omega_i^*$, $i=1,2,3$
corresponding to the 6 inherited two-cycles from the underlying $T^4$
and $\gamma_a$, $a=1,\ldots,16$ representing the 16 collapsed
spheres at the fixed points of the $\bz_2$ action. The non-trivial
part of their intersection matrix is given by
\eqn\orbbasis{\eqalign{
\omega_i\cdot\omega_j^*&=2\delta_{ij}\cr
\gamma_a\cdot\gamma_b&=-2\delta_{ab}
}.}
The basis \orbbasis, known as the Kummer basis, 
generates $H^2(K3,\bz)$ over ${\bf Q}$, but over the integers only
an index two sub-lattice of $\Gamma_{3,19}$ \ASPpoint. Observe also
that in \orbbasis\ the background $B$-field needed for a perturbative
description of the orbifold takes the form
\eqn\background{B=-{1\over{4}}\sum_{a=1}^{16}\gamma_a.}

It is important to keep in mind that the Mukai charges of a point
\eqn\point{p=Q_0\omega+Q_2+Q_4\omega^*,}
in the $\Gamma_{4,20}$ lattice are not the D$0$, D$2$ and
D$4$-brane charges as seen by a low energy observer. The latter are
shifted due to the background $B$-field \refs{\BSV\GHM}, and are given
by
\eqn\relvec{\eqalign{
q_4&=Q_4\cr
q_2&=Q_2-Q_4 B\cr
q_0&=Q_0+Q_2\cdot B-\ha Q_4B^2,
}}
where $Q_0$ already includes the shift due to the $K3$ curvature,
i.e. it is the instanton number of the bundle minus its rank.

In \RW, a $K3$ duality was formulated, which has the property that
it maps the perturbative orbifold $T^4/\bz_2$ to an orbifold of
the dual torus ${\widetilde T}^4/\bz_2$. This means in particular
that the $B$-field on the collapsed two-cycles is preserved.
The volume of the covering spaces is inverted, meaning that the
volume of the orbifold gets mapped to $1/4$ of the inverse volume.
A theory of D$0$ and D$2$-branes on the original orbifold is mapped
to the theory of D$4$-branes and other brane charges on the dual
orbifold.
Concretely, the gauge theory analysis of the theory of D$4$-branes
wrapping the dual ${\widetilde T}^4/\bz_2$ torus in \RW\ led to the
following transformation of the Kummer lattice generators and
$\omega$, $\omega^*$: 
\eqn\tdualtwo{\eqalign{
\omega_i&\mapsto\omega^*_i\hskip40pt\omega_i^*\mapsto\omega_i\hskip55pt
i=1,2,3\cr
\gamma_a&\mapsto\omega^*+\omega-\half D_a\hskip 80pt a=1,\ldots,15\cr
\gamma_{16}&\mapsto\omega^*-\omega\cr
\omega&\mapsto{\hsize=3.1cm{\line{\hfill$2\omega+2\omega^*-\half\sum_{a=1}^{16}\gamma_a$}}}\cr
\omega^*&\mapsto{\hsize=4.1cm{\line{\hfill$2\omega+2\omega^*-\half\sum_{a=1}^{15}\gamma_a+\half\gamma_{16}$}}}
}}
The asymmetry in the transformation of $\gamma_{16}$ is due to the
arbitrary assignment of $\gamma_{16}$ to the collapsed 2-cycle at the
origin. The elements $D_a$ are given by
\eqn\defined{D_{a^\prime}=\sum_{a=1}^{16}\xi_{a^\prime}^{\,a}\gamma_a,}
where according to the numbering scheme we use $\xi_{a^\prime}^{\,a}$
takes the values $1$ or $0$ according to 
\eqn\num{\xi_{a^\prime}^{\,a}=
{1\over 2}\left(1-(-1)^{4[\bv_{a^\prime}]\cdot[\bv_a]}\right)}

\subsec{Action on the branes}

In the boundary state formalism, the T-duality transformation on the
untwisted boundary states is straight-forward. It converts Dirichlet
into Neumann boundary conditions and vice versa, so that an untwisted
$|Dp;\vx\rangle$ state is mapped to an untwisted
$|D(4-p);\tilde{\vx}\rangle$ state.  Here, $\tilde{\vx}$ refers to the
dual moduli. In particular, the untwisted D$0$-brane
transforms into a state which carries only untwisted D$4$-brane
charge as seen by a low energy observer.

Since we identify the BPS vector $\omega$ with the
untwisted D$0$-brane state, the $p=4$ state should be identified with
the transform of $\omega$ under \tdualtwo. As evidence for this
identification we can compute the RR charges carried by
this BPS vector. Using \relvec, its D$0$ and D$2$-brane charges vanish
as they should. However the world volume analysis leads to twice the
fundamental D$4$-brane charge. This is because under T-duality of the
covering space, the D$0$-brane and its image transform into two
D$4$-branes on the dual covering space. The embedding of the $\bz_2$
action leads to two D$4$-branes wrapping the dual orbifold with a
non-trivial bundle structure encoded in the transformation law
\tdualtwo. From the boundary state perspective we will see that this
normalization of the D$4$-brane charge is required by the existence of
twisted D$4$-brane boundary states.

The spacetime BPS state identified with the state \twistedstates\ is
the fractional D$0$-brane of \DDG. To compare the boundary states
with the world volume analysis, the twist fields are labeled as
described before formula \tmatrixone. 
Inspection of \relvec\ shows that the
following identifications should be made:
\eqn\matchone{\eqalign{
{\hsize=3.7cm{\line{\hfill$|D0D2_{a}\rangle_+\leftrightarrow$}}}
{\hsize=2.2cm{\line{$\,\,\gamma_{a}$\hss}}}&\hskip10pt
{\hsize=2.4cm{\line{$(q_4,q_2,q_0)=(0,$}}}
{\hsize=2.7cm{\line{$\gamma_{a},\half)\hss$}}}\cr
{\hsize=3.7cm{\line{\hfill$|D0{\overline{D2}}_{a}\rangle_-\leftrightarrow$}}}
{\hsize=2.2cm{\line{$\,\,-\gamma_{a}+\omega$\hss}}}&\hskip10pt
{\hsize=2.4cm{\line{$(q_4,q_2,q_0)=(0,$}}}
{\hsize=2.7cm{\line{$-\gamma_{a},\half)\hss$}}}\cr
{\hsize=3.7cm{\line{\hfill$|{\overline{D0}}D2_{a}\rangle_-\leftrightarrow$}}}
{\hsize=2.2cm{\line{$\,\,\gamma_{a}-\omega$\hss}}}&\hskip10pt
{\hsize=2.4cm{\line{$(q_4,q_2,q_0)=(0,$}}}
{\hsize=2.7cm{\line{$\gamma_{a},-\half)\hss$}}}\cr
{\hsize=3.7cm{\line{\hfill$|{\overline{D0D2}}_{a}\rangle_+\leftrightarrow$}}}
{\hsize=2.2cm{\line{$\,\,-\gamma_{a}$\hss}}}&\hskip10pt
{\hsize=2.4cm{\line{$(q_4,q_2,q_0)=(0,$}}}
{\hsize=2.7cm{\line{$-\gamma_{a},-\half)\hss$}}}
}}
Observe that two of these D$2$-brane states also carry non-zero flux.

The Neumann state related to the Dirichlet state \orbstatetwo\ by the
duality $\ct$ is given by 
\eqn\orbstatethree{|D4D2_{a^\prime}\rangle_\pm=2^{-\half}|D4;\bv^\prime\rangle\pm2^\half|NT_{a^\prime}\dr,}
where $|NT_{a^\prime}\dr$ is the Neumann analog of \twistboseone\ built
over the $T_{a^\prime}=\ct_{a^\prime a}T_a$ twist fields. 
Obtaining \orbstatethree\ by the $\ct$ action on the fractional
D$0$-brane states means that the fixed $T^4$ Wilson line\foot{Note that
this is not a Wilson line on the $K3$ surface.} $\bv^\prime$ entering their
untwisted parts is again determined by the twist fields $T_{a^\prime}$ in the
boundary state. It is important to note though that it can be verified
independently of duality considerations that the states
\orbstatethree\ are consistent boundary states. \foot{A technical
issue alluded to before is whether other combinations of untwisted and
twisted boundary states are allowed. This applies both to the fractional
D$0$-branes and D$4$-branes. As can be verified by open string
calculations, consistency of the fractional D$0$ states with the
fractional D$4$ states excludes other combinations.}

Furthermore, we can now fix absolute normalization of D$4$-brane
charges. Being a BPS state, the twisted Neumann state \orbstatethree\
must have an integral D$4$-brane charge which we set to
unity. Comparison of this state with the inherited state in
\orbstateone\ leads to the conclusion that the inherited state indeed
carries two units of D$4$-brane charge, in agreement with the world
volume analysis. Note that the $\ct$-dual of the fractional D$0$-brane
at the origin carries the charge $\omega^*-\omega$, corresponding
to a wrapped D$4$-brane with trivial gauge bundle.

The agreement with the world volume analysis is easily checked by
comparing the action of the CFT $\ct$-transformation on the low energy
charges with the action of \tdualtwo\ on the Mukai charges.
Consider for example the BPS vector $\gamma_a$, $a\neq16$. Under
\tdualtwo\ it transforms into
\eqn\checkone{\gamma_{a^\prime}=\omega^*+\omega-\ha D_{a^\prime},}
whose RR charges are
\eqn\checktwo{\eqalign{
q_0^\prime&=0\cr
q_2^\prime&=-{1\over{4}}\left(1-(-1)^{4[\bv_{a^\prime}]\cdot[\bv_a]}\right)\gamma_a+{1\over{4}}\sum_{a=1}^{16}\gamma_a={1\over{4}}(-1)^{4[\bv_{a^\prime}]\cdot[\bv_a]}\gamma_a\cr
q_4^\prime&=1.
}}
Our CFT analysis reproduces this result. Using \tmatrixone\ and the
correct normalization for the $q_4$ charge, the charges of the $\ct$
transformed boundary states are in exact agreement with \checktwo.

Completing our discussion, we note the existence of consistent
fractional D$2$-brane boundary states which are a combination of four
twisted Ishibashi states with an untwisted D$2$-brane
state. The untwisted part corresponds to a D$2$-brane
wrapping once the fixed plane suspended by the four fixed points and
carrying half the charge of the untwisted D$2$-brane which is free to
move away from the plane. For each fixed plane there are four such
states corresponding to different relative orientations of wrapping
the collapsed $2$-cycles. It is easy to verify from the discussion
above that these states transform properly under $\ct$. More
precisely, each of the four transformed states corresponds to the 
new fixed plane intersecting the original one in one common fixed point.

\newsec{Conclusions}

When an exact CFT description is possible, boundary states are
a powerful tool in deriving results which are otherwise obtained by a
more complicated world volume analysis. In this paper we used the
boundary state formalism to derive the action \tdualtwo\ of the
inherited T-duality on the BPS states of type~II theory on
$T^4/\bz_2$. As it is not easy to find an integral basis of the
homology lattice in terms of the orbifold cycles, checking whether
\tdualtwo\ is an $O(4,20;\bz)$ element is difficult. Boundary states
are a strong evidence that \tdualtwo\ is indeed a T-duality, since the
action of the exact perturbative symmetry $\ct$ on them coincides with
that of \tdualtwo. 
The inherited T-duality can be combined with the
permutation symmetry of the four orbifold directions to yield a
sub-group of $O(4,20;\bz)$ preserving the $B$-field and square
$T^4$. It is interesting to check whether there are additional
symmetries which together with $\ct$ generates the maximal
sub-group of the T-duality group preserving this background.

{\centerline{\bf Acknowledgments}}

We would like to thank D.~E.~Diaconescu, B.~Fiol, J.~Gomis,
A.~Rajaraman, A.~Recknagel, M.~Rozali, V.~Schomerus and particularly
M.~Douglas for useful discussions. This work was supported in part by
DOE grant DE-FG02-96ER40559.

\appendix{A}{Transformation of the twist fields}

Here we show that the transformation 
\eqn\transtwist{
T_{[\bv^\prime]}=2^{-{d\over 2}}\sum_{[\bv]\in{1\over 2}W/W}
(-1)^{4\bv\cdot\bv^\prime}T_{[\bv^\prime]}}
preserves the OPE of two twist fields. Combining \transtwist\ with
\ope\ gives
\eqn\proofone{\eqalign{
T_{[\bv_1^\prime]}T_{[\bv_2^\prime]}&=
{1\over 2^d}\sum_{{[\bv_1],[\bv_2]\atop{\in\ha W/W}}}(-1)^{4\bv_1\cdot\bv_1^\prime+4\bv_2\cdot\bv^\prime_2}\hskip-15pt
\sum_{\bp\in M\atop{\bw\in
W+2\bv_1+2\bv_2}}\hskip-15pt{(-1)^{2\bp\cdot\bv_1}+(-1)^{2\bp\cdot\bv_2}\over{2}}16^{-\ha(\bp^2+\bw^2)}V_{\bp,\bw}.
}}
The first of the two terms gives
\eqn\prooftwo{\eqalign{
&{1\over 2^d}\sum_{{[\bv_1],[\bv_2]\atop{\in\ha W/W}}}
\sum_{\bp\in M\atop{\bw\in W+2\bv_1+2\bv_2}}
\ha(-1)^{2\bp\cdot\bv_1+4\bv_1\cdot\bv_1^\prime-(2\bv_1+\bw)\cdot2\bv_2^\prime}16^{-\ha(\bp^2+\bw^2)}V_{\bp,\bw}\cr
=&{1\over{2^d}}\sum_{\bp\in M\atop{\bw\in W}}\ha\sum_{[\bv_1]\in
{1\over 2}W/W}(-1)^{2\bv_1\cdot(\bp+2\bv_1^\prime-2\bv_2^\prime)-2\bw\cdot\bv_2^\prime}
16^{-\ha(\bp^2+\bw^2)}V_{\bp,\bw}\cr
=&\sum_{\bp\in2M+2\bv_1^\prime+2\bv_2^\prime\atop{\bw\in W}}\ha(-1)^{2\bw\cdot\bv_2^\prime}16^{-\ha(\bp^2+\bw^2)}V_{\bp,\bw}
}}
The second term in \proofone\ is treated in the same way and leads to 
the last line above with $\bv_1^\prime\mapsto\bv_2^\prime$. Since
$V_{\bp,\bw}$ is the vertex operator $V^\prime_{\bw,\bp}$ of the
$\ct$-transformed theory, this shows that the OPE \ope\ is preserved.

\listrefs

\end